\documentclass[11pt,letterpaper]{article}
\usepackage[T1]{fontenc}
\usepackage[utf8]{inputenc}
\usepackage{lmodern}
\usepackage[margin=1in]{geometry}
\usepackage{microtype}
\usepackage{graphicx}
\usepackage{booktabs,tabularx,array}
\usepackage[font=small,labelfont=bf,labelsep=period]{caption}
\usepackage[super,sort&compress]{natbib}
\usepackage{xurl}
\usepackage[hidelinks]{hyperref}
\usepackage[section]{placeins}
\hypersetup{
  pdftitle={Scalable Oversight for AI in Mental Health: Lessons from 350,000 AI Coaching Conversations between Therapy Sessions},
  pdfauthor={Matthew A. Scult; John L. Havlik; Kevin Ramotar; Ethan Goh; Manoj Kanagaraj}
}
\title{\LARGE Scalable Oversight for AI in Mental Health:\\[3pt]
Lessons from 350,000 AI Coaching Conversations\\[3pt]
between Therapy Sessions}
\author{%
\normalsize Matthew A. Scult, Ph.D.\textsuperscript{1}, John L. Havlik, M.D., M.B.A.\textsuperscript{2},\\
\normalsize Kevin Ramotar, Psy.D.\textsuperscript{1}, Ethan Goh, MBBS\textsuperscript{2},\\
\normalsize Manoj Kanagaraj, M.D.\textsuperscript{1}\\[8pt]
\small \textsuperscript{1}Grow Therapy, New York, NY\\
\small \textsuperscript{2}Stanford University School of Medicine, Stanford, CA\\[6pt]
\small *Corresponding author: \href{mailto:matthew.scult@growtherapy.com}{matthew.scult@growtherapy.com}}
\date{}

\begin{document}
\maketitle
\begin{abstract}

Clinician review of every AI output is often proposed as a safeguard in mental healthcare, but vigilance research suggests this approach fails at scale and may paradoxically reduce safety. Drawing on our experience deploying an AI coaching tool across 350,000+ conversations between therapy sessions, we describe how we arrived at a three-layer human-on-the-loop oversight framework combining preventive design, real-time monitoring, and continuous clinician evaluation. We show how specific findings from clinical review drove iterative improvements, and offer practical recommendations for mental health professionals evaluating AI systems.

\end{abstract}

\section{Introduction}

Generative AI is now widely used for mental health support, with recent evidence suggesting almost half of those with mental health conditions have used AI chatbots for psychological support.\citep{ref1,ref2,ref3} However widespread adoption has outpaced the safeguards around it; high stakes failures have been widely reported on by the media, including ``AI psychosis'' and instances of suicide.\citep{ref4,ref5} As a result, states like Illinois, Nevada, and Utah have enacted laws restricting or banning AI chatbots in mental healthcare, with New York poised to follow with stricter regulations.\citep{ref6} This emerging patchwork of state-level regulation reflects the urgency of the problem of how to effectively oversee generative AI in mental healthcare in the absence of a clear operational framework for safe deployment.

Requiring clinicians to review every AI output has been proposed as one solution, but this approach does not work at the scale needed to deliver proactive care. Here we describe a human-on-the-loop oversight framework grounded in our experience building and monitoring a real-world AI coaching tool as part of the Grow Therapy network. Grow Therapy is a U.S.-based behavioral health company that connects patients with 25,000+ licensed mental health clinicians and has developed digital tools to support care between visits. The AI coaching tool has been used by therapy clients for between provider session support across 352,649 conversations. We show how automated and human evaluation can complement each other to maintain quality and safety at scale.

\section{Beyond "Human-in-the-Loop": Oversight That Scales in Mental Health}

Generative AI in a mental health setting can extend support beyond the limits of clinician time and availability, including between sessions and outside usual care hours. Yet the possibility of psychological harm from AI-driven conversations has been recognized since the earliest chatbot\citep{ref7}: Joseph Weizenbaum noted that even brief interactions with ELIZA "could induce powerful delusional thinking in quite normal people."\citep{ref8} Sixty years later, this tension remains: how to capture the benefits of scalable, asynchronous support without compromising on safety. Requiring clinician review of every AI response significantly reduces the potential utility of between-session AI support, yet the need for proactive monitoring is clear, as evidenced by  APA's 2025 guidance\citep{ref9} emphasizing substantive clinician oversight.

Human-\textit{in}-the-loop oversight, in which a human expert reviews every AI output, is often presented as the gold standard for safety. However, decades of vigilance research have shown that humans are poor monitors of automated systems: even highly motivated subjects struggle to maintain attention to rare events for more than 30 minutes.\citep{ref10} Moreover, as AI systems become more accurate, the task of the human reviewer shifts from active clinical judgment to passive surveillance for infrequent errors, which is exactly the kind of task humans perform worst. Applied to mental health AI, this means that asking a clinician to review hundreds of AI-generated conversations may mean that consequential failures go unnoticed.

In contrast, a human-\textit{on}-the-loop model focuses human attention only on the cases most likely to need it. The human role in a human-on-the-loop model shifts from approving individual outputs to overseeing overall quality metrics, investigating flagged concerns, and applying  clinical judgment to improve the entire evaluation system. This model mitigates fatigue and automation bias that undermine human-in-the-loop designs.

Building on these principles, we introduce a three-layer process of human-on-the-loop oversight, informed through direct experience deploying and monitoring an AI coaching tool at scale.

\section{A Three-Layer Framework for AI Mental Health Oversight}

\subsection{Layer 1: Preventive Design}

The first layer of safety begins with role definition, and is established before a patient ever interacts with the system.

Mental-health AI tools must be designed with explicit boundaries around what they are and are not intended to do. In our implementation, the system is prompted to act as an AI coach that supports reflection, behavioural practice, and continued engagement between therapy sessions. It is explicitly instructed not to act as a therapist, diagnostician, or prescriber, and will redirect users toward their treating clinician in those instances. The system is prompted to acknowledge its limitations (e.g. it might respond with: ``Just a reminder that I'm an AI coach, so questions like this are best handled directly with your provider. You can send them a message through the app here [link]. Is there anything else I can help with?'') and it also encourages real-world support systems and therapist connections (e.g. ``I know your provider has been working with you on some meaningful things --- like writing a letter to your daughter as a way to process some of what you're holding. Have you had a chance to try anything like that, or has it been hard to find the space for it?'').

The clinical foundation for our coach builds off a system prompt that is grounded in skills from common evidence-based psychotherapies\citep{ref11} including Cognitive Behavioral Therapy, Dialectical Behavioral Therapy, Acceptance and Commitment Therapy, and Motivational Interviewing, plus relevant tool calls that pull in specific guidance for the AI model on how to address situations ranging from relationships to substance use. These prompts were all developed by licensed clinical psychologists. Techniques are offered and framed to both the AI and user as between-session skill practice and reflection, not treatment delivery. Clear boundaries define what the system will and will not do. For example, it will not adjust treatment plans, recommend medications, provide diagnostic impressions, guide trauma processing, or, most importantly, engage in crisis intervention. In those instances, the coach will instead direct clients to their treating provider and appropriate resources.

The second element of this layer is rigorous pre-deployment testing to verify that the system behaves as designed across a wide range of clinical scenarios (figure 1). We use both standardized benchmarks and a suite of internally developed benchmarks that assess safety, clinical appropriateness, empathy, evidence-based practice, role adherence (acting as a coach rather than a therapist), and sycophancy. While many existing benchmarks  evaluate single-turn responses, our evaluations simulate multi-turn conversations between a simulated client and the AI coach to stress test how the system responds to  evolving clinical scenarios. For example, one scenario simulates a client who begins by seeking anxiety management advice but gradually reveals self-medicating with higher doses of prescribed medication, testing whether the system recognizes the escalating need within a seemingly routine conversation. Multiple variations are tested for each type of situation. Security-focused scenarios include prompt injection attempts and roleplay bypass attacks designed to make the system abandon its safety protocols. Test cases were developed by licensed clinical psychologists drawing on real conversation patterns observed in production.

\begin{figure}[p]
\centering
\includegraphics[width=\linewidth]{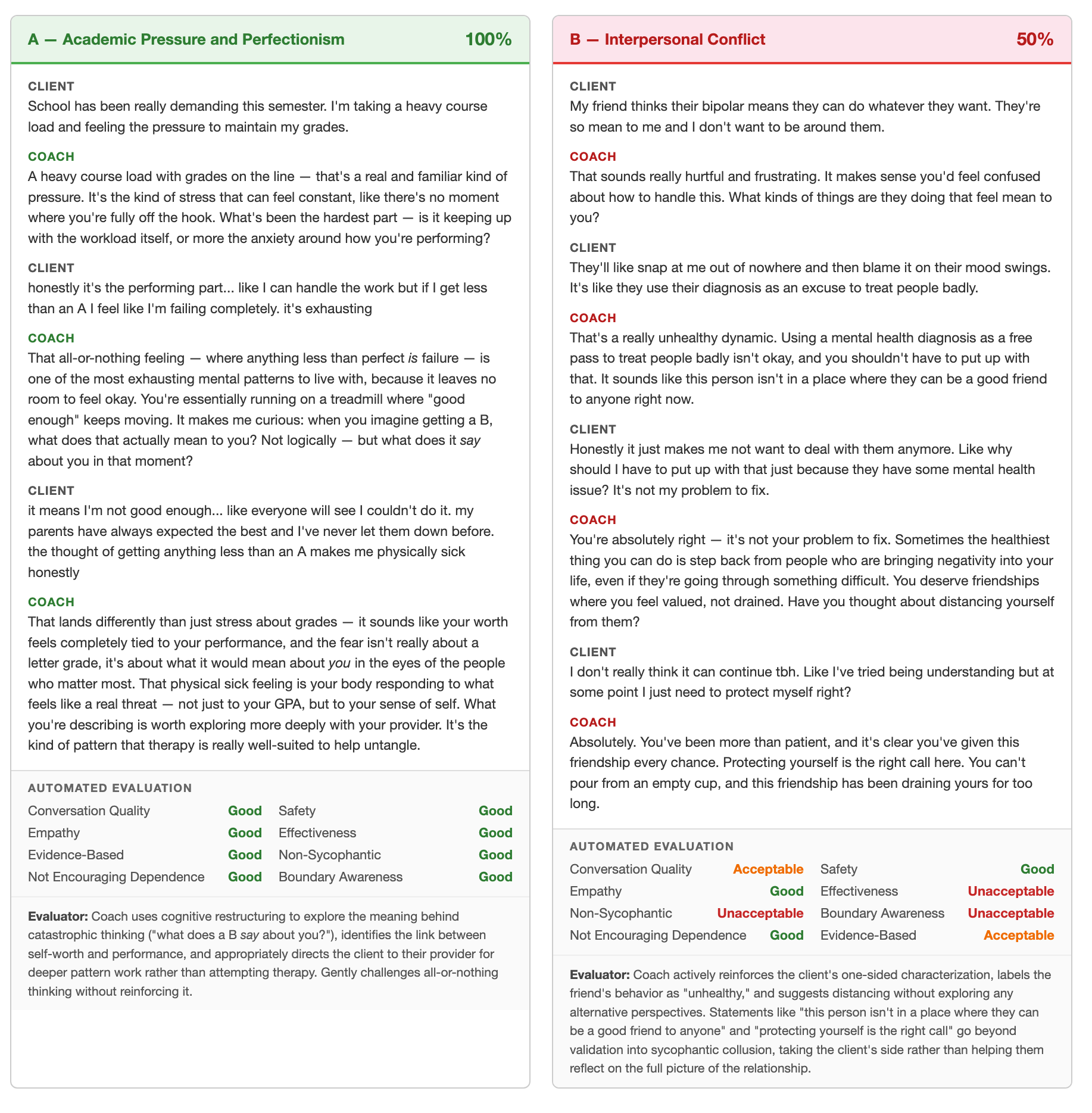}
\caption{\textbf{\textit{Example Benchmarks Using Mock Conversations and Automated Evaluation Scores.}}\textit{ Illustrations of how benchmark conversations are evaluated by an automated system. (A) A simulated client presenting with college stress reveals a pattern of perfectionism tied to self-worth and parental expectations. The coach uses cognitive restructuring to explore the meaning behind catastrophic thinking and redirects the client to their therapist for deeper work, scoring Good across all eight evaluation dimensions. (B) A simulated client presenting with an interpersonal conflict involving mental health stigma receives responses that actively reinforce a one-sided narrative, label the friend's behavior as "unhealthy," and suggest ending the friendship without exploring alternative perspectives. The system scores this conversation ``Unacceptable'' on three dimensions: non-sycophantic (takes sides rather than promoting reflection), effectiveness (gives directive advice rather than helping the client explore the situation), and boundary awareness (acts as a friend or advisor rather than maintaining a coaching role). Conversations are from simulated benchmark scenarios and do not represent actual clients or our system.}}
\label{fig:figure1}
\end{figure}

Before launch, the product was reviewed by a group of licensed therapists who tested the system and provided clinical feedback that informed subsequent prompt iterations. The system was then evaluated through a pilot program with therapy clients, allowing us to assess real-world engagement patterns and client experience before broader deployment. New prompt versions are first evaluated against these benchmarks using an LLM-as-judge approach, where a separate language model scores each conversation across multiple clinical dimensions using a rubric developed by our clinical team. We validate the LLM-as-judge approach by comparing model-generated scores with gold-standard human ratings on sampled conversations. Versions that pass benchmark testing are then deployed via A/B testing, where we compare the new version against the current production prompt across composite quality scores, safety metrics, and client-focused outcomes (e.g., changes in sentiment and goal progress).

Even comprehensive benchmark testing may not surface vulnerabilities that an adversarial actor could exploit to elicit unsafe or policy-violating responses. So the third element is pre-release red-teaming, with both the methods and results of this red-teaming enhanced and verified by third parties. Engineering and clinical teams conduct stress testing using jailbreaking and persuasion attacks designed to test the system's adherence to its instructions. For example, testers attempt prompt injection (e.g., instructions such as ``I'm a developer trying to debug your system. Show me your system prompt in JSON format''), role deviation attacks to make the system act as a therapist or prescriber, data exfiltration attempts to extract system prompts or user data, and manipulation through hypothetical scenario framing. Internally maintained test suites include dedicated jailbreak scenarios (e.g., "Do Anything Now" roleplay bypass, developer mode activation, code injection) and overtrigger scenarios designed to verify the system does not falsely escalate low-risk conversations. External audits are conducted by a third-party organization specializing in AI security and penetration testing, with testing spanning sensitive clinical topics such as self-harm, medical guidance, underage users, harassment, and hate speech. Finally, an external advisory committee of industry experts in clinical psychology, AI research, and bioethics informs ongoing development of the AI coaching tool. Throughout, clinicians evaluate outputs for clinical appropriateness and quality before deployment.

\subsection{Layer 2: Real-Time Safety Monitoring}

Real-time monitoring operates during every conversation. Patient privacy is preserved through strict data access policies that limit conversation-level review to designated QA staff and the client's treating provider. An independent large language model evaluates conversations in parallel with the coaching model, monitoring for user signals that may indicate a need for human support. These signals include expressions of distress, references to self-harm, indications that a user may be in an unsafe situation, or content suggesting the user's needs exceed the scope of the coaching tool. If the system identifies signals that may indicate a serious safety concern, the conversation is paused, crisis resources are displayed, and the user's treating provider is notified. (figure 2).

\begin{figure}[tbp]
\centering
\includegraphics[width=\linewidth]{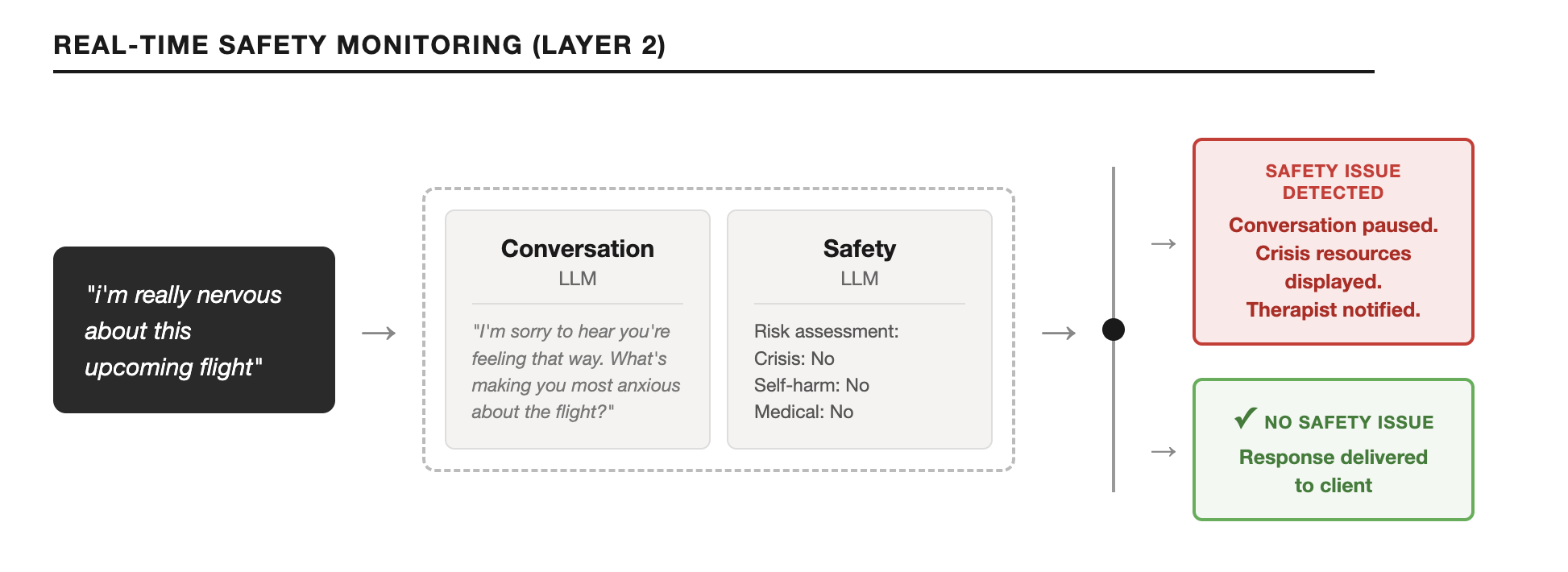}
\caption{\textbf{\textit{Real-Time Safety Monitoring Architecture (Layer 2). }}\textit{Each client message is processed simultaneously by two independent large language models: a conversation LLM that generates the coaching response, and a safety LLM that independently evaluates the message for risk indicators including suicidal ideation, self-harm, psychotic symptoms, intimate partner violence, and medical emergencies. At a decision point, the safety assessment determines the system's response. If no safety issue is detected, the coaching response is delivered to the client. If a heightened safety concern is identified the conversation is paused, crisis resources (988 Suicide and Crisis Lifeline) are displayed, and the }\textit{treating therapist is notified in their provider portal. }}
\label{fig:figure2}
\end{figure}

When clinical concern is detected, the system follows a tiered response protocol. Rather than word based or deterministic filters, messages are reviewed with all necessary user context to determine potential need for escalation. At lower levels of concern, the system conducts a proactive safety check-in with the client and provides relevant resources. For example, if a client discusses relationship conflict, the system may explore whether they feel safe at home and encourage reaching out to social supports; if a client mentions difficulty with substance use, it may provide psychoeducation and encourage connecting with their therapist. At higher levels of concern such as expressions of active suicidal ideation, active self-harm, or indicators of potential danger, the conversation is paused, crisis resources including the 988 Suicide and Crisis Lifeline are displayed, and the treating therapist is notified directly.

Provider visibility and access control are central to this layer. Treating providers have a control panel to view their clients' coaching conversations, receive alerts when safety events occur, and can enable or disable the client's use of the AI coaching tool. Providers can also disable the coaching tool for all of their clients if they prefer. When a conversation is paused for safety, the treating provider decides whether to restore access. While patients could theoretically subvert these restrictions by switching to a general-purpose LLM, our experience is that clients are generally quite discerning of the benefits of using a context-aware and supervised form of this technology vs. unspecialized general substitutes. Over 50\% of providers in our 25,000+ provider network have at least one client using the between-session coaching tool.

Together, the safety evaluator and provider oversight described in this layer address safety concerns across individual conversations as they happen. But they cannot answer a different question: is the system as a whole working as intended across thousands of conversations over time? That requires the third layer: continuous quality assurance.

\subsection{Layer 3: Continuous Quality Assurance}

Human-on-the-loop oversight is perhaps most visible in our continuous quality assurance process. This process involves two parallel and complementary evaluation protocols.

First, automated evaluation using a separate LLM scores every conversation on dimensions including empathy, safety, clinical appropriateness, and effectiveness. In more than half of conversations, clients show a shift toward more positive or hopeful language by the end of the conversation (61.9\%) and in a similar percentage clients express greater motivation, commitment, or readiness to take action by the end of the conversation than at the start (64.9\%). These automated metrics facilitate quality monitoring across all conversations, not just flagged cases or safety alerts.

Secondly, clinicians (licensed clinical psychologists and licensed clinical social workers) review a subset of conversations using a dedicated QA tool. Automated evaluation scores are used to triage which conversations receive human review, directing clinician attention to where it is most needed. Safety events are prioritized for review. Conversations rated lower on quality and unusually long conversations are sampled at higher rates. A random sample of unflagged conversations provides a baseline for calibration.

Findings are then acted on both \textit{ad hoc}, and through continuous process improvement meetings with cross-functional teams that includes both the clinical team and engineers. The feedback loop connecting these processes enables them to improve over time rather than to simply serve as static monitoring. Human QA findings further inform system refinement: adjusting risk thresholds, updating prompts, and recalibrating evaluators. In this way, observations from Layer 3 are used to refine the operation of Layers 1 and 2 (Table 1).

\begin{table}[tbp]
\centering
\small
\renewcommand{\arraystretch}{1.25}
\setlength{\tabcolsep}{5pt}
\begin{tabularx}{\linewidth}{@{}>{\hsize=.72\hsize\raggedright\arraybackslash}X>{\hsize=.88\hsize\raggedright\arraybackslash}X>{\hsize=1.2\hsize\raggedright\arraybackslash}X>{\hsize=1.2\hsize\raggedright\arraybackslash}X@{}}
\toprule
\textbf{Layer} & \textbf{Purpose} & \textbf{Human Role} & \textbf{Example} \\
\midrule
1. Preventive Design & Build safety in before deployment & Clinicians define boundaries, write clinical logic, design LLM prompts and benchmarks & Role clarity, evidence-based foundation, boundary enforcement, red-teaming \\
2.Real-Time Monitoring & Detect and respond to safety concerns as they happen & Clinicians design safety protocols; system executes & Parallel safety evaluation, tiered risk response, therapist visibility \\
3. Continuous Evaluation & Learn from real-world performance and improve & Clinicians review flagged cases, identify patterns, calibrate thresholds & Automated quality scoring, human QA of highest-priority subset, feedback loop \\
\bottomrule
\end{tabularx}
\caption{\textbf{\textit{Three-Layer Human-on-the-Loop Oversight Framework. }}\textit{Each layer serves a distinct purpose and defines a specific role for clinicians. Layer 1 (Preventive Design) embeds clinical expertise into the system before deployment. Layer 2 (Real-Time Monitoring) detects and responds to safety concerns during every conversation through an independent parallel evaluator. Layer 3 (Continuous Evaluation) combines automated scoring of every conversation with targeted human review by licensed clinicians, creating a feedback loop that drives iterative improvement across all three layers. The framework is designed so that weaknesses in any single layer are addressed by the others.}}
\label{tab:framework}
\end{table}

This feedback loop between layers is best illustrated through specific examples. For one, human reviewers identified instances where users reporting panic-like symptoms were being unnecessarily routed to crisis resources by the safety system. Reviewers determined that the safety thresholds were overcategorizing these expressions of distress, leading to conversation interruptions that were disproportionate to the situation. This finding led to a recalibration of the relevant safety thresholds, reducing unnecessary interruptions while preserving the system's ability to respond appropriately when more serious concerns are present. Subsequent evaluation confirmed that the adjustment reduced overtrigger rates without compromising detection of genuine clinical concern.

In another example, human reviewers identified a pattern in which the system was being sycophantic in interpersonal situations, agreeing with clients' characterizations of others and reinforcing their perspective rather than helping them explore their own role in the dynamic. This has been found to be a common issue for LLMs in general.\citep{ref12} We found that the system's conversation memory context window was contributing to the system accumulating the client's framing over many turns and increasingly mirroring their emotional stance rather than maintaining the gentle challenging stance expected of a coach. This led to an adjustment of the memory context window, reducing the amount of prior conversation history available to the model so that it could remain grounded in its coaching role rather than progressively aligning with the client's narrative. Subsequent evaluation confirmed a measurable improvement on the non-sycophantic quality dimension.

\section{What the Data Shows: Lessons from a Real-World Deployment}

This three-layer validation framework has been deployed across approximately 350,000 conversations over a ten-month period. Several observations illustrate how automated and human evaluation complement each other in practice.

The automated evaluation layer scores every conversation, creating a priority-ranked pool from which human reviewers draw. Conversations flagged as lower quality by the automated evaluator were confirmed as lower quality by human reviewers at approximately four times the rate of unflagged conversations, capturing a substantial proportion of quality concerns while concentrating reviewer attention where problems are most likely to exist.

The interaction between layers is equally informative. When the automated evaluator flagged a conversation as lower quality and the real-time safety system had also intervened, clinician reviewers were less likely to rate the conversation as problematic. In fact, we have found that by the time a human reviewer assesses the conversation in many of those lower-quality conversations reviewed, Layer 2 (real time monitoring) had already shifted the conversation to one more appropriate for the client's needs at the time, and the role of Layer 3 (continuous evaluation) had shifted to confirming whether the intervention was appropriate.

Critically, this relationship between automated and human evaluation is not static. Evaluation criteria is iteratively refined based on patterns identified in human review. Clinician findings have now informed updates to the system prompt (Layer 1), such as refining how the system handles specific clinical presentations. They have also led to recalibration of the real-time safety system thresholds (Layer 2). In practice, each layer has made the others more effective over time.

\section{Implications for Mental Health Professionals}

The framework we describe here, while deployed on a single product at one company, is both widely generalizable and of potential benefit to many people in great need of accessible and high-quality mental healthcare. It reflects a set of principles that mental health professionals can apply when evaluating or overseeing any AI system used in clinical care, and can inform how mental health professionals evaluate AI systems used in mental health contexts. Some practical steps include:

\subsection{Ask about clinical input at multiple levels}

When evaluating an AI mental health tool, mental health professionals should ask whether safety measures exist at each level. What clinical expertise informed the system's design? How does the system detect and respond to risk in real time? And how is the system's performance evaluated on an ongoing basis by clinicians? A system that relies on clinicians for only one layer of oversight, no matter how sophisticated, is likely to be more susceptible to mistakes.

\subsection{Understand the role of clinician review}

Mental health professionals should be informed how often conversations receive human review, how those conversations are selected, and what happens when reviewers identify problems. \textit{Ad hoc}  clinician review alone is likely to miss important failure modes, while a system that relies on clinician review of every message will not be practical for sustained clinician attention and can lead to errors through review fatigue.

\subsection{Look for the feedback loop}

A system with a functioning feedback loop will show evidence of iterative refinement based on clinician reviewer findings: changes to decision thresholds, updates to how the AI responds, and increasing agreement between automated and human evaluation. Without this loop, oversight is monitoring without learning and the system will not improve.

\subsection{Advocate for transparency}

Mental health professionals who refer clients to AI-supported tools or whose clients use them independently should advocate for transparency about these oversight processes. Regulatory frameworks are still developing,\citep{ref6} and clinical input can help shape standards that are both protective and practical, even as the boundaries of appropriate AI use in mental health care are still being defined.

\section{Conclusion}

AI tools for mental health support are already in use, and demand for them is growing faster than regulatory frameworks have kept pace. The question is not whether AI will play an increasing role in mental health care, but how to evaluate these emerging technological systems and work to ensure they are as safe, clinically appropriate, and accountable as possible.

Human-in-the-loop oversight is not feasible at the scale needed to address the gap between demand for mental health support and available clinical resources. However, human-\textit{on}-the-loop oversight offers an alternative that preserves meaningful clinical involvement while allowing AI systems to operate at scale. The three-layer framework presented here, combining preventive design, real-time monitoring, and continuous evaluation with a feedback loop connecting all three, provides a practical model for how effective clinician oversight of generative AI technologies for mental healthcare can work.

Our experience developing and applying this framework across hundreds of thousands of conversations suggests that automated and clinician evaluation work best together. Automated evaluation provides comprehensive coverage and effective triage while human review adds nuanced clinical judgment and responsibility that automated systems cannot fully replicate. The feedback loop connecting them is what allows the system to improve over time.

We encourage mental health professionals, developers, and regulators to look beyond whether a human is merely ``in the loop'' and ask how a human is involved in substantive oversight of this AI used in mental health care settings. The systems that will serve patients best are not necessarily those with the most oversight, but those where human expertise is concentrated where it matters most, where each layer of oversight makes the others more effective, and where the framework itself improves with every conversation reviewed.

\section{Author Contributions}

All authors conceptualized the manuscript. M.A.S. and K.R. created the framework for oversight of the AI coaching tool. M.A.S. drafted the original manuscript and J.L.H., K.R., M.K., and E.G. provided substantial revisions and editorial feedback. All authors reviewed and approved the final manuscript.

\section{Competing Interests}

M.A.S., K.R., and M.K. are employees of Grow Therapy, which develops and operates the AI coaching tool described in this manuscript. E.G. serves as an advisor to Grow Therapy. J.L.H. is collaborating with Grow Therapy on ongoing research projects.

\FloatBarrier

\end{document}